\documentclass{article}

\usepackage{arxiv}

\usepackage[utf8]{inputenc} 
\usepackage[T1]{fontenc}    
\usepackage{hyperref}       
\usepackage{url}            
\usepackage{booktabs}       
\usepackage{amsfonts}       
\usepackage{nicefrac}       
\usepackage{microtype}      
\usepackage{lipsum}		
\usepackage{graphicx}
\usepackage{natbib}
\usepackage{doi}
\usepackage{float}
\usepackage{amsmath}
\usepackage{bbm}
\usepackage{dsfont}

\title{Generalizing Impermanent Loss on Decentralized Exchanges with Constant Function Market Makers}

\author{ Rohan Tangri\\
	\texttt{rohan@compasslabs.ai} \\
	\And
	Peter Yatsyshin\thanks{Peter Yatsyshin is also a Research Associate at Imperial College London, United Kingdom (p.yatsyshin@imperial.ac.uk)} \\
	\texttt{p.yatsyshin@compasslabs.ai} \\
	\And
	Elisabeth A. Duijnstee \\
	\texttt{elisabeth@compasslabs.ai} \\
	\AND
	Danilo Mandic \\
	\texttt{d.mandic@imperial.ac.uk} \\
}

\renewcommand{\headeright}{}
\renewcommand{\undertitle}{}
\renewcommand{\shorttitle}{Tangri, Yatsyshin, Duijnstee and Mandic}

\hypersetup{
pdftitle={The Liqudity Providing Problem},
pdfsubject={q-bio.NC, q-bio.QM},
pdfauthor={David S.~Hippocampus, Elias D.~Striatum},
pdfkeywords={Liquditiy Provisioning},
}

\begin{document}
\maketitle

\begin{abstract}
    Liquidity providers are essential for the function of decentralized exchanges to ensure liquidity takers can be guaranteed a counterparty for their trades. However, liquidity providers investing in liquidity pools face many risks, the most prominent of which is impermanent loss. Currently, analysis of this metric is difficult to conduct due to different market maker algorithms, fee structures and concentrated liquidity dynamics across the various exchanges. To this end, we provide a framework to generalize impermanent loss for multiple asset pools obeying any constant function market maker with optional concentrated liquidity. We also discuss how pool fees fit into the framework, and identify the condition for which liquidity provisioning becomes profitable when earnings from trading fees exceed impermanent loss. Finally, we demonstrate the utility and generalizability of this framework with simulations in BalancerV2 and UniswapV3.
\end{abstract}


\section{Introduction}

Decentralized finance (DeFi) aims to provide a trustless system of finance by replacing third-parties who control liquidity flows with smart contracts that manage rules for issuing debt and trading. Between 2019 and 2021, the industry has grown to manage over \$150 billion of assets. The largest applications within the DeFi ecosystem are decentralized exchanges (DEXs), which coordinate large-scale trading of digital assets in a non-custodial manner. This is achieved through automated algorithms, as opposed to a centralized exchange (CEX) which requires an intermediary to facilitate the transfer and custody of funds.

There are two types of agents that can interact with a DEX, liquidity providers (LPs) and liquidity takers (LTs). LPs provide their digital assets to a liquidity pool which acts as the counterparty for LTs. For every trade by an LT, the pool takes a portion of the trade as a fee, which is then distributed amongst the LPs and protocol treasury. In this way, LPs are motivated to participate with the incentive of generating a passive income in a way not accessible on CEXs. The more liquidity provided to a liquidity pool, the better the trade dynamics for LTs (e.g. lower slippage), therefore attracting more trade activity. Therefore, a big effort is made by DEXs to attract LPs to their platforms, typically in the form of financial incentives such as liquidity mining.

A major barrier for many LPs is impermanent loss (IL). IL is the opportunity cost of providing assets in a liquidity pool instead of holding those same assets over the investment period. \cite{permanent-loss} have shown that for convex constant product market makers, $\text{IL} \ge 0$. Therefore, liquidity provisioning is only profitable when trading and liquidity mining fees compensate for this loss. Research by \cite{https://doi.org/10.48550/arxiv.2111.09192} revealed that over 50\% of LPs on UniswapV3 suffer IL which is not compensated by fee earnings. They also showed that the most popular strategy is to passively invest, with 30\% of LPs holding their positions for longer than a month. Furthermore, each DEX can operate with a different market maker algorithm, fee structure and (concentrated) liquidity dynamics, and the current absence of any formal general framework makes it difficult to easily analyze IL across exchanges. This work aims to establish the missing general framework and employs it to consider different forms of IL over UniswapV3 and BalancerV2.

\section{Fundamentals of Constant Function Market Maker Pools}

Consider a pool of $N$ assets. Each asset is characterized by its quantity in the pool, given by the number of the respective token, and its internal instantaneous price with respect to the other assets in the pool. Denote the quantities of the assets in the pool as $Q = [q_1, \dots, q_N]^T$ and the internal instantaneous price of asset $i$ with respect to asset $j$ as $Z_{i,j}$, meaning one unit of asset $i$ is exchangeable for $Z_{i,j}$ units of asset $j$.

There are two events that can trigger a change in pool state: a trade and a quote. A trade occurs when an LT exchanges assets with the liquidity pool. Unlike a swap which is constrained to an exchange of two assets (\cite{https://doi.org/10.48550/arxiv.2210.01227}), a trade allows for an exchange of up to $N$ assets. Suppose that the LT wants to add and/or remove assets from the pool in exchange for asset $k$, then a trade can be defined as

\begin{equation}
    X = [\Delta q_1, \dots, \Delta q_k, \dots, \Delta q_N]^T, \quad \prod_{i=1}^N\Delta q_i < 0
    \label{eq:trade}
\end{equation}
where $\Delta q_i$ represents the change in the $i$-th asset quantity from the perspective of the pool. The quantites $\Delta q_i, i \ne k$ are given by the trader, and only $\Delta q_k$ needs to be solved for by the pool.

A quote occurs when an LP provides or withdraws liquidity from the liquidity pool. A quote can be defined as

\begin{equation}
\begin{gathered}
    Y = [\Delta q_1, \dots, \Delta q_N]^T, \quad \begin{cases} \forall_i \Delta q_i > 0 & \text{if providing} \\
    \forall_i \Delta q_i < 0 & \text{if withdrawing}\end{cases} \\
    \frac{\Delta q_i}{\Delta q_j} = \frac{q_i}{q_j}, \quad i \ne j
    \label{eq:liquidity-event}
\end{gathered}
\end{equation}

At all times, the $N$ assets of a pool satisfy the automated market maker (AMM), which dictates new asset quantities given an event. The most common type of AMM implemented in DEXs is the constant function market maker (CFMM) of the form 

\begin{equation}
F(q_1, \dots, q_N; \zeta) = K
\label{eq:cfmm}
\end{equation}
where $F: \mathrm{R}_+^N\to \mathrm{R}_+$ is a continuous map, $K$ is the depth of the pool and $\zeta$ are any further parameters.

The instantaneous price of asset $i$ with respect to asset $j$ is defined by the quantity of asset $j$ that can be traded for an infinitesimally small quantity of asset $i$. For a trade where one of $\Delta q_i$ and $\Delta q_j$ is negative and the other is positive, the instantaneous price is given by

\begin{equation}
\begin{gathered}
    Z_{i, j} = \lim_{\Delta q_i \rightarrow 0} - \frac{\Delta q_j}{\Delta q_i} = - \frac{\partial q_j}{\partial q_i} \\
    Z_{i, i} = 1
    \label{eq:z}
\end{gathered}
\end{equation}

where the relationship between $q_i$ and $q_j$ is derived via the CFMM formula in (\ref{eq:cfmm}). Assuming that $Z_{i,j} \ge 0$, then the CFMM is constrained by the condition $\frac{\partial q_j}{\partial q_i} \le 0$. Furthermore, given the pool should always sell at a higher price than it buys, the CFMM should be convex with respect to the asset quantities (\cite{permanent-loss}).

Finally, each asset has an associated weight, $w_i$, which is defined as the proportion of value in the pool held in the $i$-th asset, that is

\begin{equation}
    w_i = \frac{q_i Z_{i,j}}{\sum_{i=1}^N q_i Z_{i,j}}
\end{equation}

\section{Uniform Pool Mechanics}

In a uniform liquidity pool, LPs provide liquidity over the entire price range $[0, + \infty ]$ for LTs to make trades with. An LP therefore only needs to decide which pools to provide liquidity for based on the assets held within. An example of the dynamics described here can be found in Section \ref{sec:BalancerV2}.

\subsection{Solving an Event}
\label{sec:AMM}

For simplicity, this section will ignore fees. When a trade occurs, the CFMM equation in (\ref{eq:cfmm}) can be solved to obtain the final asset quantities. Given a trade, the amount of asset $k$ to be deposited or withdrawn, $\Delta q_k$, needs to solved for: 

\begin{equation}
\begin{gathered}
    K = F(q_1, \dots, q_N; \zeta) \\
    F(q_1 + \Delta q_1, \dots, q_k + \Delta q_k, \dots, q_N + \Delta q_N; \zeta) = K
    \label{eq:amm-trade}
\end{gathered}
\end{equation}

On the other hand, when a quote occurs, a new depth parameter, $K_Y$, needs to be calculated from the new asset quantities; the new depth calculation is given by

\begin{equation}
    K_Y = F(q_1 + \Delta q_1, \dots, q_N + \Delta q_N; \zeta)
    \label{eq:amm-le}
\end{equation}

\subsection{Pool Share}

Each LP has a share of the pool they are invested in. This is tracked by LP tokens, where each LP owns a proportion of these tokens and they are created and destroyed to affect the total supply as LPs provide or withdraw liquidity. Let $\Tilde{q}_i^j$ be the quantity of asset $i$ owned by the $j$-th LP. The pool share of the $j$-th LP, $\Tilde{r}_j$, is defined by asset quantities, not prices, and is given by

\begin{equation}
    \Tilde{r}_j = \frac{\Tilde{q}_i^j}{q_i} = \frac{\sum_{i=1}^N \Tilde{q}_i^j}{\sum_{i=1}^N q_i}
\end{equation}

\subsection{Pool Fees}
\label{sec:fees}

For every trade with a pool, a percentage of the inbound assets to the pool is taken as a fee and distributed amongst the LPs and protocol treasury. Denote $\gamma$ by the total fee paid by LTs and $\phi$ by the protocol fee, that is, the percentage of the total fee designated to the protocol treasury. The trade quantities for the $i$-th asset including fees, $\Delta q_i^\gamma$, can then be derived from the trade quantities excluding fees, $\Delta q_i$, which are solved for in (\ref{eq:amm-trade}), to give

\begin{equation}
    \Delta q_i^\gamma = \begin{cases}
    \frac{\Delta q_i}{1-\gamma}, \quad \Delta q_i > 0 \\
    \Delta q_i, \quad \Delta q_i \le 0
    \end{cases}
\end{equation}

Given a trade including fees, the quantities of inbound assets can be defined as

\begin{equation}
    X_+ = [\max(\Delta q_1^\gamma, 0), \dots, \max(\Delta q_N^\gamma, 0)]^T
\end{equation}

and the total fees for all LPs generated from the trade, $\lambda$, is given by

\begin{equation}
    \lambda = \gamma (1-\phi) X_+^T \in \mathbb{R}^N
\end{equation}

These fees are generally reinvested into the LP position, although in some cases they can an also be kept in a separate account. Either way, the LP receives their fees on top of their underlying position in the pool once they exit their position. Furthermore, the $j$-th LP specific fees, $\Tilde{\lambda}_j$, can be derived from the total fees generated by the pool and the LP pool share as

\begin{equation}
    \Tilde{\lambda}_j = \Tilde{r}_j \lambda
    \label{eq:specific-lp}
\end{equation}

Assuming trading fees are collected in the pool, the CFMM dynamics are affected in the event of a trade since a new pool depth, $K_X$, needs to be computed for the new trade quantities, given as

\begin{equation}
    K_X = F(q_1^t + \Delta q_1^\gamma, \dots, q_N^t + \Delta q_N^\gamma; \zeta)
\end{equation}

\begin{figure}[H]
    \centering
    \includegraphics[width=0.5\textwidth]{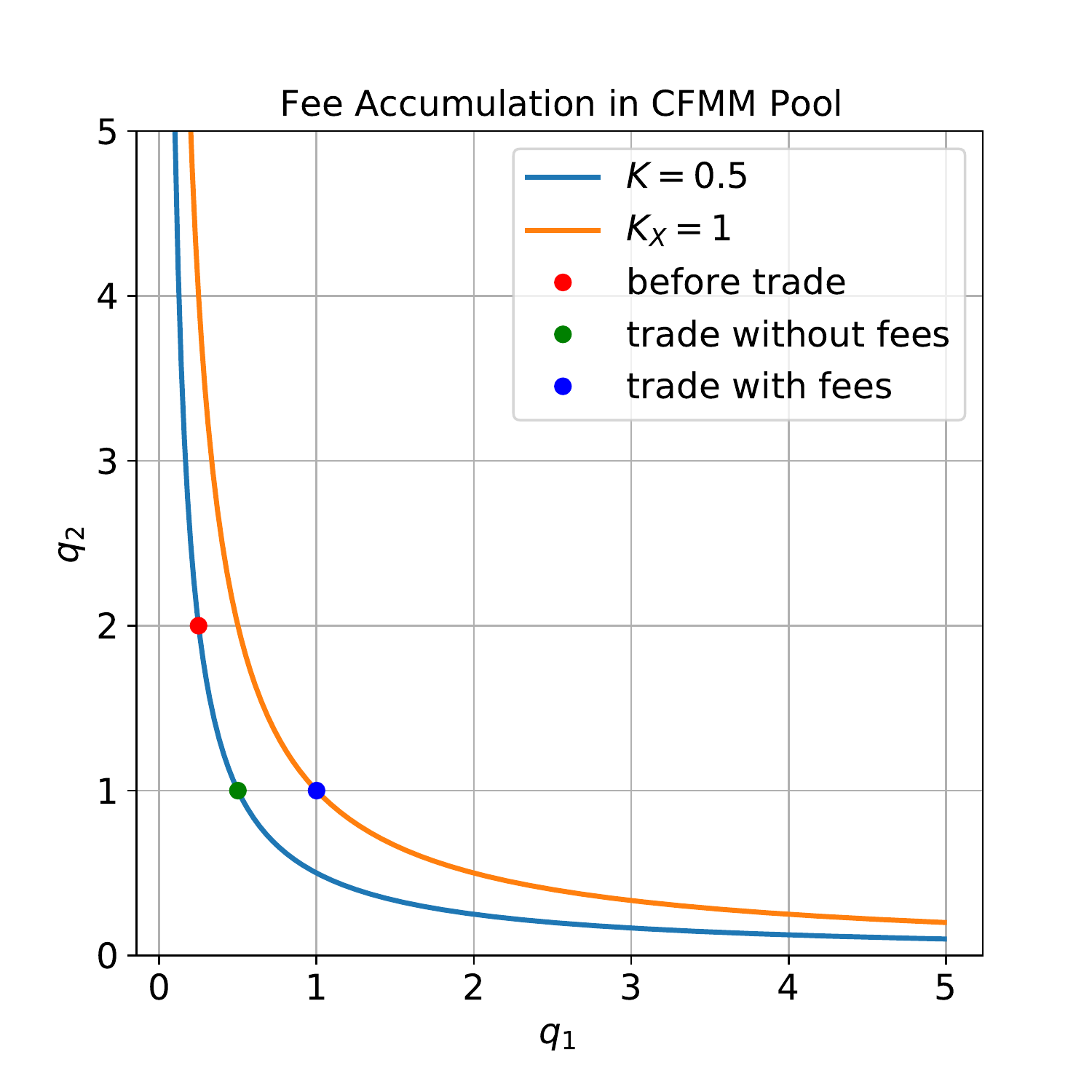}
    \caption{This figure shows an example CFMM with fee accumulation in a two asset pool. The trade fees increase the depth of the pool, $K_X > K$, since the fees are only added to the incoming assets to the pool.}
    \label{fig:fee-accumulation}
\end{figure}

Liquidity mining is a feature in many DEXs, and allows LPs to stake their LP tokens for extra rewards; generally, this is used to incentivize LPs to keep their positions for longer time periods. Each DEX handles this differently, but these rewards are also collected once an LP exits their position; in the meantime, these are generally kept in a separate account.

\section{Concentrated Pool Mechanics}

Concentrated liquidity (CL) allows LPs to specify an instantaneous price range for the liquidity they want to provide. In practice, the continuous price space is discretized into ticks such that LPs can provide liquidity in a range between any two ticks which need not be adjacent. An example of the dynamics described here can be found in Section \ref{sec:UniswapV3}.

\subsection{Price Surface}

Given a pool with $N$ assets, an LP must provide liquidity in any instantaneous price range, $R_{l,u}$, for an asset pair $(n, n+1)$, with bounds $(Z_{n, n+1}^l, Z_{n, n+1}^u]$ for $N-1$ pairs:

\begin{equation}
    R_{l,u} =
    \left[ {\begin{array}{cc}
    Z_{1,2}^l & Z_{1,2}^u \\
    \vdots & \vdots \\
    Z_{N-1,N}^l & Z_{N-1,N}^u \\
  \end{array} } \right] \in \mathbb{R}^{N-1 \times 2}
\end{equation}

From the perspective of the pool, the price ticks form a price surface, that is, a grid in $N-1$ dimensions with $T_d$ ticks in the $d$-th dimension.

A unit price range, $R_{l,l+1}$, defines a price range bounded by adjacent ticks such that the space cannot be subdivided further given a price surface

\begin{gather}
    R_{l,l+1} =
    \left[ {\begin{array}{cc}
    Z_{1,2}^l & Z_{1,2}^{l + 1} \\
    \vdots & \vdots \\
    Z_{N-1,N}^l & Z_{N-1,N}^{l + 1} \\
  \end{array} } \right] \in \mathbb{R}^{N-1 \times 2}
\end{gather}

The active price range defines the unit price range in which the current instantaneous prices reside such that at any time

\begin{equation}
    R_A = \{ R_{l,l+1} \} =
    \left[ {\begin{array}{cc}
    Z_{1,2}^l & Z_{1,2}^{l + 1} \\
    \vdots & \vdots \\
    Z_{N-1,N}^l & Z_{N-1,N}^{l + 1} \\
  \end{array} } \right] \forall_{n \in \{ 1, \dots, N-1 \}} Z_{n, n+1}^l < Z_{n, n+1}^t \le Z_{n, n+1}^{l + 1}
  \label{eq:active-range}
\end{equation}

\subsection{Virtual Pool}

Denote $\Tilde{q}_i^j$ by the quantity of asset $i$ provided by the $j$-th LP in the price range $R_j$ and let each unit price range in the pool $R_{l, l+1}$ have $i$-th asset quantity $q_i^{l,l+1}$. Given $L$ liquidity providers in the pool, the asset quantities in a unit price range can be defined as a transformation of the LP quantities

\begin{equation}
    q_i^{l, l+1} = \sum_{j=1}^L \Tilde{q}_i^j \mathds{1}_{R_{l,l+1} \subseteq R_j}
    \label{eq:unit-quantities}
\end{equation}
where $\mathds{1}_A$ is the indicator function of interval $A$.

In practice, these real asset quantities are not used to define pool behavior, but instead, a virtual pool with virtual quantities and virtual depth is defined over the active price range (\cite{uniswapv3}). Specifically, the active price range only needs to hold enough assets to cover an instantaneous price movement to a tick boundary. To achieve this, the CFMM specified in (\ref{eq:cfmm}) is translated such that the position is solvent exactly within the active range. A virtual pool can be defined over any finite price range. Denote $q_i^{l,u}$ by the real quantity of the $i$-th asset over the price range $R_{l,u}$, $q_i^v$ by the virtual quantity of the $i$-th asset over the same range, and $K_v$ by the virtual depth. Then

\begin{gather}
    F(q_1^v, \dots, q_N^v; \zeta) = K_v \\
    q_i^v = g_i(q_i^{l,u})
\end{gather}

\subsection{Solving an Event}

Given a new quote, the new quantities in each unit price range can be computed as in (\ref{eq:unit-quantities}).

Given a trade, $X$, if the active range does not change, the CFMM equation looks very similar to (\ref{eq:amm-trade}), just using the fixed virtual depth and virtual asset quantities to solve for $\Delta q_k$, that is

\begin{equation}
\begin{gathered}
    K_v = F(q_1^v, \dots, q_N^v; \zeta) \\
    F(q_1^v + \Delta q_1, \dots, q_k^v + \Delta q_k, \dots, q_N^v + \Delta q_N; \zeta) = K_v
    \label{eq:cl-trade}
\end{gathered}
\end{equation}

However, suppose a trade causes the instantaneous rate to cross $M - 1$ separate price ticks, then the order is executed as $M$ separate transactions with the respective values for the virtual depths. A tick is crossed once the real quantities of any asset in the unit range are depleted. However, the boundary condition can also be calculated using the virtual pool in the active range. Define $\beta \in [0, 1]$ by the control, the fraction of a trade that has been completed taking into account all quantity changes the trader has control over. In other words, given a trade needs to be solved for a quantity $\Delta q_k$, $\beta$ is applied as a fraction to the trade for $i \ne k$ as

\begin{equation}
    X_\beta = [\beta \Delta q_1, \dots, \Delta q_k(\beta), \dots, \beta \Delta q_N]^T, \quad 0 \le \beta \le 1
    \label{eq:beta-trade}
\end{equation}
where $\Delta q_k(\beta)$ can be solved for by the CFMM equation
\begin{equation}
    F(q_1^v + \beta \Delta q_1, \dots, q_k^v + \Delta q_k(\beta), \dots, q_N^v + \beta \Delta q_N; \zeta) = K_v
\end{equation}

Therefore, the instantaneous price at some time $\Delta t$ into the trade, $Z_{i,j}^{t+\Delta t}$, can be represented as a function of the control $\beta$, in the form

\begin{gather}
    Z_{i,j}^{t+\Delta t}(\beta) = 
    \begin{cases}
        - \frac{\partial q_j}{\partial q_i} \Bigr|_{\substack{q_i=q_i^v+\beta \Delta q_i \\ q_j=q_j^v+\beta \Delta q_j}} \quad i,j \ne k \\ 
        - \frac{\partial q_j}{\partial q_i} \Bigr|_{\substack{q_i=q_i^v+\beta \Delta q_i \\ q_j=q_j^v+\Delta q_j(\beta)}} \quad j=k \\
        - \frac{\partial q_j}{\partial q_i} \Bigr|_{\substack{q_i=q_i^v+\Delta q_i(\beta) \\ q_j=q_j^v+\beta \Delta q_j}} \quad i=k \\
    \end{cases}
\end{gather}

and so given some active price range, the expression can be solved to find $\beta_{n,n+1}^l$ and $\beta_{n,n+1}^u$, that is, the controls that push the instantaneous rate of asset pair $n, n+1$ to the lower and upper bound of the active range.

\begin{equation}
\begin{gathered}
    Z_{n,n+1}^{t+\Delta t}(\beta_{n,n+1}^l) = Z_{n,n+1}^l \\
    Z_{n,n+1}^{t+\Delta t}(\beta_{n,n+1}^u) = Z_{n,n+1}^{l+1}
    \label{eq:cross-condi}
\end{gathered}
\end{equation}

\subsection{Pool Share}

In pools with CL, LPs own a share of the liquidity in each unit price range. Similarly to uniform pools, this is defined by asset quantities. Given a unit price range $R_{l,l+1}$, the $j$-th LP with liquidity in the price range $R_j$ owns

\begin{equation}
    \Tilde{r}_j^{l,l+1} = \frac{\Tilde{q}_i^j}{q_i^{l,l+1}} \mathds{1}_{R_{l,l+1} \subseteq R_j} = \frac{\sum_{i=1}^N \Tilde{q}_i^j}{\sum_{i=1}^N q_i^{l,l+1}} \mathds{1}_{R_{l,l+1} \subseteq R_j}
\end{equation}

\subsection{Pool Fees}

Given a trade over $M$ active price ranges $R_A = \{ R_{l,l+1}^1, \dots, R_{l,l+1}^M \}$ and controls $\{ \beta_1, \dots, \beta_M \}$ such that $\sum_{i=1}^M \beta_i = 1$, the incoming assets to the pool over each active range is

\begin{equation}
    X_+^m = [\max(\beta_m \Delta q_1, 0), \dots, \max(\Delta q_k(\beta_m), 0), \dots, \max(\beta_m \Delta q_N, 0)]^T
\end{equation}

so that the pool fee over each active range are

\begin{equation}
    \lambda_m = \gamma (1-\phi) X_+^m
\end{equation}

The total fees paid by LTs are distributed amongst the LPs who own liquidity in the active ranges of the trade. Therefore, the $j$-th LP will receive

\begin{equation}
    \Tilde{\lambda}_j = \sum_{m=1}^M \Tilde{r}_j^m \lambda_m
\end{equation}

In CFMMs with CL, fee income is not automatically reinvested in the pool because LPs can provide liquidity in various ranges simultaneously, so the distribution and reinvestment of fees across different ranges is a choice made by the LP. This is in contrast to the fee structure for CFMMs without CL as described in Section \ref{sec:fees} where fees are more commonly automatically compounded into the LP position (not including liquidity mining fees).

\section{Arbitrage}
\label{sec:LR}

Let the external FIAT prices of the assets in a pool be represented by $P = [p_1, \dots, p_N]$. A difference between the external FIAT price and the internal instantaneous price of assets in a pool leads to arbitrage trades, which alter the quantities of assets. The trades continue until the new instantaneous prices reflect the current FIAT prices. This process is referred to as \textbf{equilibration} of the pool. In an equilibrated pool, the instantaneous prices match the external FIAT prices of the assets. This price balance condition (PBC) can be encoded by $N-1$ equations, in the form

\begin{equation}
    Z_{i,j} = \frac{p_i}{p_j}, \quad i \neq j
    \label{eq:lr}
\end{equation}

If equation (\ref{eq:lr}) does not hold, the pool is not in equilibrium and an arbitrage opportunity exists which LTs can take advantage of. Let a specific arbitrage, $\delta_{i,j}$ refer to the arbitrage opportunity of an asset $i$ with respect to an asset $j$, that is

\begin{equation}
    \delta_{i,j} = Z_{i,j} - \frac{p_i}{p_j}
    \label{eq:specific-arb}
\end{equation}

Further, let arbitrage, $\delta_j$ be a reduction of the specific arbitrage such that it defines the total arbitrage opportunity in asset $j$ given by

\begin{equation}
    \delta_j = \sum_{i=1}^N Z_{i,j} - \frac{p_i}{p_j}
    \label{eq:arb}
\end{equation}

\section{Impermanent Loss} 

Impermanent loss (IL) is the difference between the value of assets provided as liquidity in a pool and the value of those same assets had the LPs simply held them instead of providing liquidity. Given a pool with $N$ assets, let there be $k$ trades, $\{ X_1, \dots, X_k \}$, and $l$ quotes, $\{ Y_1, \dots, Y_l \}$, in the time range $[t, \dots, T]$. Further, denote $Y_{[t, \dots, T]}$ by the sum of all quotes in the time range as

\begin{equation}
    Y_{[t, \dots, T]} = \sum_{i=1}^l Y_i = [\Delta q_1^Y, \dots, \Delta q_N^Y]
\end{equation}

The strict definition of IL for a pool ignoring fees is given in a FIAT currency by

\begin{equation}
    \text{IL}_p = \sum_{i=1}^{N}(q_i^t + \Delta q_i^Y) p_i^T - \sum_{i=1}^{N}q_i^T p_i^T
\end{equation}

However, it can also be given in terms of an asset $j$ as

\begin{equation}
    \text{IL}_j = \sum_{i=1}^{N}(q_i^t + \Delta q_i^Y) Z_{i,j}^T - \sum_{i=1}^{N}q_i^T Z_{i,j}^T
    \label{eq:IL-j}
\end{equation}

%
%
%
%

\subsection{Relative Value}

Relative value (RV) is a measure that has a more intuitive meaning than IL, and is defined as the ratio between the value of assets provided as liquidity and the value of those same assets had the LP simply held them instead of providing liquidity. Again, there are two ways to define RV depending on the frame of reference: either with respect to external FIAT prices or internal instantaneous prices. The RV with respect to the external FIAT price ignoring fees is defined as

\begin{gather}
    V_p = \frac{\sum_{i=1}^{N}q_i^T p_i^T}{\sum_{i=1}^{N}(q_i^t + \Delta q_i^Y) p_i^T} 
    \label{eq:rv-p}
\end{gather}

and the RV with respect to asset $j$ in the pool ignoring fees is defined as

\begin{equation}
    V_j = \frac{\sum_{i=1}^{N}q_i^T Z_{i,j}^T}{\sum_{i=1}^{N}(q_i^t + \Delta q_i^Y) Z_{i,j}^T} 
    \label{eq:rv-z}
\end{equation}

If $V_p = V_j$, then the pool is in equilibrium, else there is an arbitrage opportunity that can be exploited by LTs.

A change in RV (and IL) only arises when there is a trade in the pool. Assume there are no trades and only quotes in the time range $[t, \dots, T]$, then

\begin{equation}
\begin{split}
    V_j = V_p &= \frac{\sum_{i=1}^{N}q_i^T p_i^T}{\sum_{i=1}^{N}(q_i^t + \Delta q_i^Y) p_i^T} \\
    &= \frac{\sum_{i=1}^{N}q_i^T p_i^T}{\sum_{i=1}^{N}q_i^T p_i^T} \\
    &= 1
\end{split}
\label{eq:rv-y}
\end{equation}

\subsection{Fee-Adjusted Relative Value}

Fee-adjusted relative value (FARV) measures the relative value of a pool taking into account the fees generated by trades that are allocated to LPs. For clarity, denote $q_i$ by the quantity of the $i$-th asset in the pool without fees, the set $\{\lambda_1, \dots, \lambda_k\}$ by the fees generated from each of the $k$ trades, and $\lambda_{[t, \dots, T]}$ by the sum of the fees as

\begin{equation}
    \lambda_{[t, \dots, T]} = \sum_{i=1}^k \lambda_i = [\lambda_1^T, \dots, \lambda_N^T]
\end{equation}

The FARV with respect to FIAT prices and with respect to asset $j$ are defined as

\begin{gather}
    V_p^\lambda = \frac{\sum_{i=1}^{N}(q_i^T + \lambda_i^T) p_i^T}{\sum_{i=1}^{N}(q_i^t + \Delta q_i^Y) p_i^T} \label{eq:farv-p} \\
    V_j^\lambda = \frac{\sum_{i=1}^{N}(q_i^T + \lambda_i^T) Z_{i,j}^T}{\sum_{i=1}^{N}(q_i^t + \Delta q_i^Y) Z_{i,j}^T} \label{eq:farv-z}
\end{gather}

In the event of a quote, FARV behaves in the same way as RV since no fees are collected, so that $V_p^\lambda = V_j^\lambda = 1$.

\subsection{Liquidity Provider Profitability Condition}

For a pool with a CFMM as defined in (\ref{eq:cfmm}), $\text{IL} \ge 0$ due to the convexity of the CFMM (\cite{permanent-loss}). The LP profitability condition defines the space where the value of fees earned by an LP outweigh the value of IL suffered. For pools without CL, IL maps linearly from the pool to an individual LP via the pool share ratio. The LP profitability condition is given for pools without CL as

\begin{equation}
    V_j^\lambda \ge 1
\end{equation}

This is not the case for pools with CL because the IL of the pool does not map linearly with the IL of an individual LP. Therefore, the FARV of the individual LP, $\Tilde{V}_j^\lambda$, should be more explicitly considered. Consider a time range $[t, \dots, T]$, denote $\Tilde{q}_i^t$ by the quantity of asset $i$ owned by a single LP and $\Tilde{q}_i^T$ by the final quantity of asset $i$ owned by a single LP ignoring fees. Further, let $\Delta \Tilde{q}_i^Y$ denote the total quote of asset $i$ given by the LP and $\Tilde{\lambda}_i^T$ the total fees allocated to the LP in asset $i$ over the time period. Then the LP profitability condition can be defined for a pool with CL as

\begin{equation}
    \Tilde{V}_j^\lambda = \frac{\sum_{i=1}^{N}(\Tilde{q}_i^T + \Tilde{\lambda}_i^T) Z_{i,j}^T}{\sum_{i=1}^{N}(\Tilde{q}_i^t + \Delta \Tilde{q}_i^Y) Z_{i,j}^T} \ge 1
\end{equation}

\section{Examples}

This section demonstrates the utility of the framework discussed here over different DEXs.

\subsection{UniswapV3}
\label{sec:UniswapV3}

At the time of writing, UniswapV3 was the second largest DEX by liquidity with \$3.49b TVL and \$4.11b trade volume (7 day). It features concentrated liquidity and has a common type of CFMM known as a constant product market maker (CPMM), constrained to two assets with time-invariant weights $w_1=w_2=0.5$, that is

\begin{equation}
    q_1 q_2 = K
\end{equation}

From this market maker, the instantaneous price can be derived as

\begin{equation}
\begin{split}
    Z_{1,2} = - \frac{\partial q_2}{\partial q_1} =\frac{K}{q_1^2} = \frac{q_1 q_2}{q_1^2} = \frac{q_2}{q_1}
\end{split}
\end{equation}

\begin{figure}[H]
    \centering
    \includegraphics[width=0.5\textwidth]{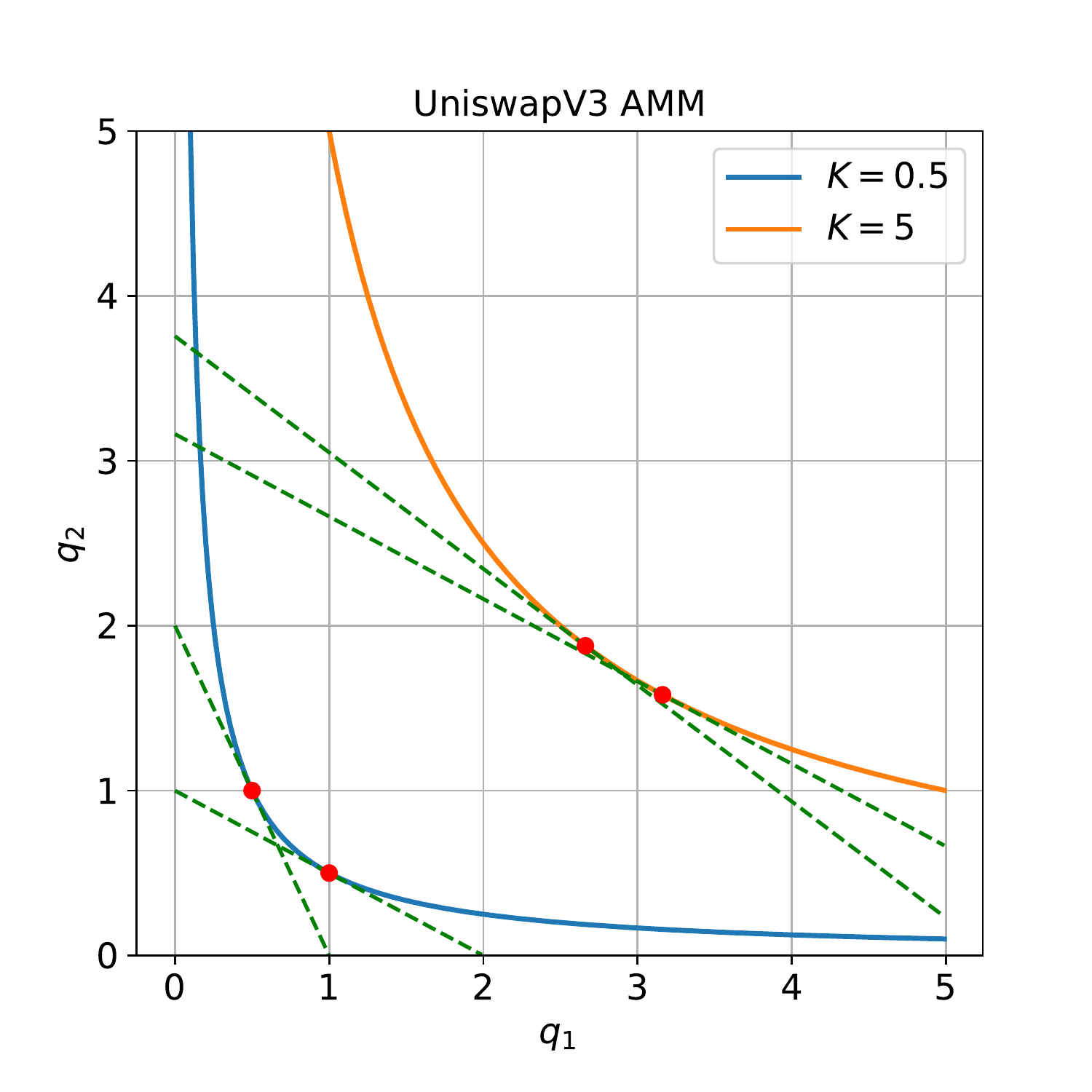}
    \caption{The CPMM is designed so that liquidity in the pool can never be exhausted. The blue and orange lines show the CPMM with different depths, and the green lines show the tangents to both AMM curves at the red dots given a starting price $Z_{1, 2}=0.5$ and a trade $X=[-0.5, \Delta q_2]^T$. The negative gradient of the tangent to the curve defines the instantaneous price. As asset $q_1$ is depleted, the gradient grows to infinity, and so the unit cost of $q_1$ also grows. Given the same trade and starting instantaneous price, a pool with greater depth will have a reduced price impact. This can be observed as the gradient of the tangent lines in the pool with depth $K=5$ changes less than the gradient of the tangent lines in the pool with depth $K=0.5$.}
    \label{fig:uniswap-amm}
\end{figure}

Further, the virtual pool with virtual quantities $q_i^v$ and virtual depth $K_v$ in the price range $(Z_{1,2}^l, Z_{1,2}^u]$ with real quantities $q_i^{l,u}$ is defined as (\cite{uniswapv3})

\begin{gather}
    q_1^v = q_1^{l,u} + \sqrt{\frac{K_v}{Z_{1,2}^l}}\\
    q_2^v = q_2^{l,u} + \sqrt{K_v Z_{1,2}^u}\\
    q_1^v q_2^v = K_v
\end{gather}

\begin{figure}[H]
    \centering
    \includegraphics[width=\textwidth]{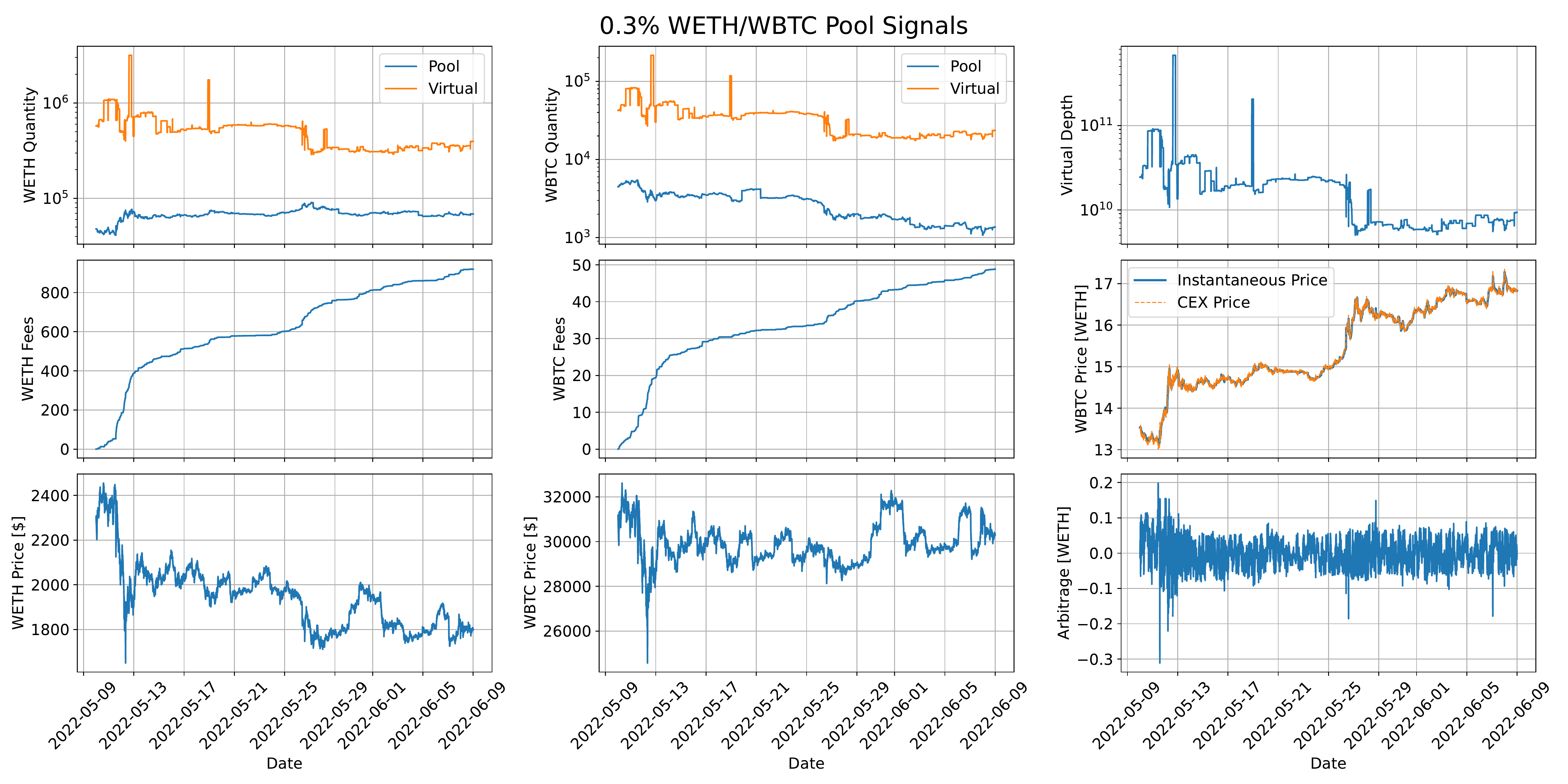}
    \caption{Virtual quantities and depths are given for the active price range. The High correlation between WBTC and WETH virtual quantities ($\rho = 0.99609592$) indicates that the majority of activity in the pool is explained by quotes rather than trades.}
    \label{fig:uni-assets}
\end{figure}

\begin{figure}[H]
    \centering
    \includegraphics[width=\textwidth]{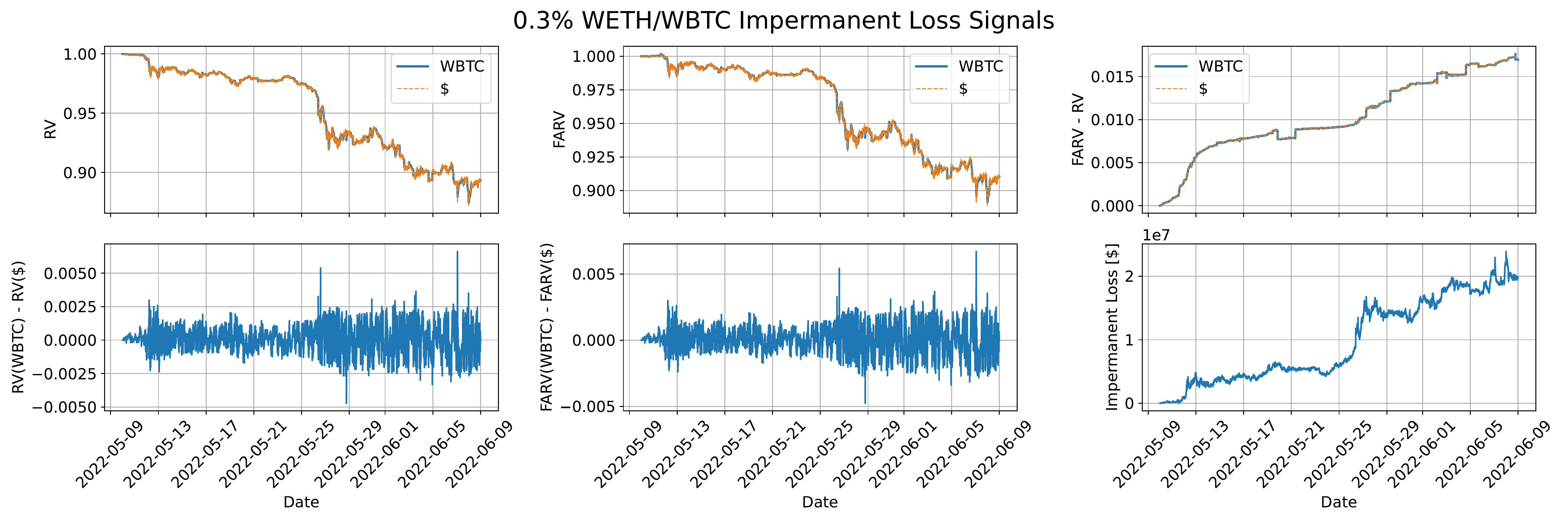}
    \caption{IL is always positive because of the convexity of the AMM function (\cite{permanent-loss}), and FARV shows that the pool as a whole would have been roughly 10\% wealthier if the assets were kept separate from the pool. FARV and RV currencies do not represent the unit of the signal; rather, they represent the currency with which the signal was generated.}
    \label{fig:uni-il}
\end{figure}

In UniswapV3, the pool fees are fixed to a discrete set such that each asset pair has a unique pool for each fee tier. The figures in this section analyze pool address \href{https://etherscan.io/address/0xCBCdF9626bC03E24f779434178A73a0B4bad62eD}{\texttt{0xCBCdF9626bC03E24f779434178A73a0B4bad62eD}} over a month with a sampling period of one minute. This pool contains WETH and WBTC with a 0.3\% pool fee. Figure \ref{fig:uni-assets} shows the evolution of the quantities of assets in the pool, and the fees generated from trades and arbitrage, while Figure \ref{fig:uni-il} shows how the variants of IL change over time.

\subsection{BalancerV2}
\label{sec:BalancerV2}

At the time of writing, BalancerV2 was one of the top 5 DEX by liquidity with \$1.2b TVL and \$963m trade volume (7 day). It does not feature concentrated liquidity and has a generalized constant product market maker for $N$ assets known as a constant mean market maker (CMMM) given by (\cite{https://doi.org/10.48550/arxiv.1911.03380})

\begin{equation}
    \prod_{i=1}^N (q_i) ^ {w_i} = K
\end{equation}

From this market maker the instantaneous price can be derived as (\cite{balancerv2})

\begin{equation}
    Z_{i,j} = \frac{q_j w_i}{q_i w_j}
\end{equation}

\begin{figure}[H]
    \centering
    \includegraphics[width=0.5\textwidth]{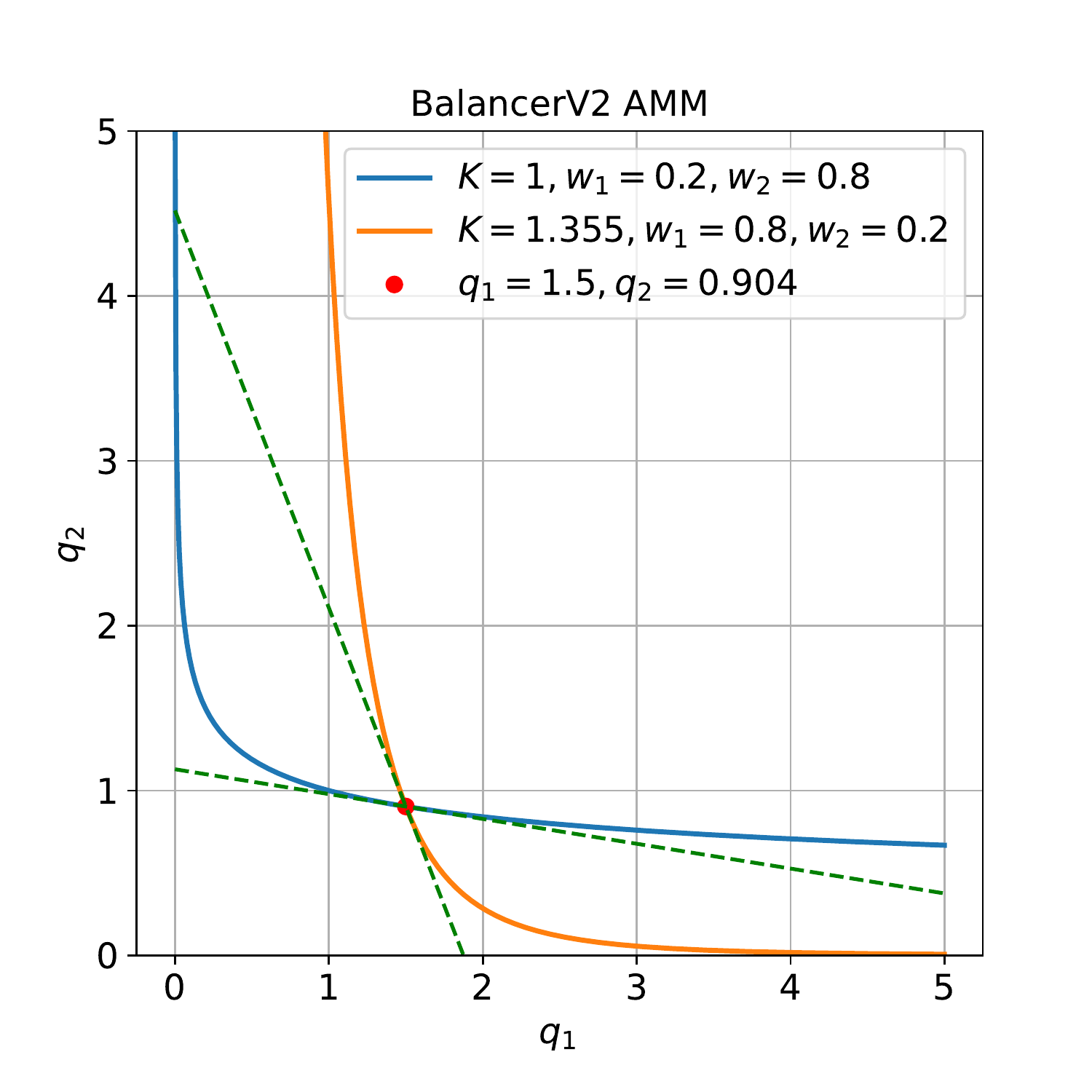}
    \caption{The orange and blue lines show the BalancerV2 AMM curves for different asset weights. Given the asset quantities in a pool remain constant, as shown by the red dot, a change in the weights changes the pool depth and the overall AMM curve, which leads to a change in the instantaneous price. The green lines show the tangents to both AMM curves at the fixed asset quantities. The change in the gradient of the tangents reflects the change in the instantaneous price.}
    \label{fig:bal-amm}
\end{figure}

For weighted math pools (pools that contain a non-stablecoin asset), the weights are set at pool creation and are time-invariant. Therefore, given the AMM and pool fee percentage are static, the LP profitability condition is used to calculate whether the LP fees overcome IL for any trade in isolation. In BalancerV2, a trade is constrained between 2 assets, given by asset $x$ and asset $y$, and fees accumulate in the pool. Denote $X=[\Delta q_1, \dots, \Delta q_N]$ by the trade without fees, $\Gamma_i$ by the total fees of asset $i$ collected by the pool, and $\lambda_i$ by fees of asset $i$ allocated to the LPs in the pool as

\begin{gather}
    \Gamma_i = \gamma \max\bigg(\frac{\Delta q_i}{1-\gamma}, 0\bigg) =
    \begin{cases}
        \frac{\gamma \Delta q_i}{1-\gamma}, \quad \Delta q_i > 0 \\
        0, \quad \Delta q_i \le 0
    \end{cases} \\
    \lambda_i = (1-\phi) \Gamma_i =
    \begin{cases}
        \frac{\gamma (1-\phi) \Delta q_i}{1-\gamma}, \quad \Delta q_i > 0 \\
        0, \quad \Delta q_i \le 0
    \end{cases}
\end{gather}
where $\gamma$ and $\phi$ are defined as in Section \ref{sec:fees}.

There are three pool asset quantities to consider: $q_i^t$ represents the starting quantity of asset $i$ before the trade, $q_i^t + \Delta q_i$ represents the quantity of asset $i$ after the trade without fees, and $q_i^t + \Delta q_i + \Gamma_i$ represents the quantity of asset $i$ after the trade with fees. The final instantaneous price of asset $i$ with respect to asset $j$ after the trade with fees is denoted $\hat{Z}_{i,j}^T$, with the hat symbol to emphasize its existence on a new AMM curve as

\begin{equation}
    \hat{Z}_{i,j}^T = \frac{(q_j^t + \Delta q_j + \Gamma_j)w_i}{(q_i^t + \Delta q_i + \Gamma_i)w_j}
\end{equation}

These parameters can be used to solve the LP profitability condition for $\Delta q_x$ and $\Delta q_y$ to identify the space of trades that are profitable for LPs to participate in, as given by

\begin{equation}
    \begin{gathered}
        V_y^\lambda = \frac{\sum_{i=1}^{N}(q_i^t + \Delta q_i + \lambda_i) \hat{Z}_{i,y}^T}{\sum_{i=1}^{N}(q_i^t + \Delta q_i^Y) \hat{Z}_{i,y}^T} \ge 1 \\
        \sum_{i=1}^{N}(q_i^t + \Delta q_i + \lambda_i) \hat{Z}_{i,y}^T - \sum_{i=1}^{N}q_i^t \hat{Z}_{i,y}^T \ge 0 \\
        (q_x^t + \Delta q_x + \lambda_x) \hat{Z}_{x,y}^T + (q_y^t + \Delta q_y + \lambda_y) - q_x^t \hat{Z}_{x,y}^T - q_y^t \ge 0 \\
        (\Delta q_x + \lambda_x) \hat{Z}_{x,y}^T + \Delta q_y + \lambda_y \ge 0
    \end{gathered}
\end{equation}

The AMM itself can be used to define the second equation to solve for the two unknowns (\cite{balancerv2})

\begin{equation}
    \begin{gathered}
        (q_x^t)^{w_x} (q_y^t)^{w_y} = (q_x^t + \Delta q_x)^{w_x}(q_y^t + \Delta q_y)^{w_y} \\
        \Delta q_y = q_y^t\bigg[ \bigg(\frac{q_x^t}{q_x^t + \Delta q_x}\bigg)^\frac{w_x}{w_y} - 1\bigg]
    \end{gathered}
\end{equation}

Now, suppose the trade deposits asset $x$ to the pool in exchange for asset $y$ such that $\Delta q_x > 0$ and $\Delta q_y < 0$, then

\begin{equation}
    \Delta q_x w_x q_y^t (q_x^t)^{\frac{w_x}{w_y}}\bigg(1 + \frac{\gamma(1-\phi)}{1-\gamma}\bigg) + q_y^t w_y\bigg(q_x^t + \Delta q_x + \frac{\gamma \Delta q_x}{1-\gamma}\bigg) \bigg((q_x^t)^\frac{w_x}{w_y} - (q_x^t + \Delta q_x)^\frac{w_x}{w_y}\bigg) \ge 0
    \label{eq:balancer-lp-profitable}
\end{equation}

The figures in this section analyze pool address \href{https://etherscan.io/address/0xa6f548df93de924d73be7d25dc02554c6bd66db5}{\texttt{0xA6F548DF93de924d73be7D25dC02554c6bD66dB5}} over a month with a sampling period of one minute. This pool is a weighted math pool on BalancerV2 containing WETH and WBTC with weights $w_{BTC} = w_{ETH} = 0.5$ and fees $\gamma = 0.0025, \phi=0.1$. Given these pool parameters, (\ref{eq:balancer-lp-profitable}) can be reduced as

\begin{equation}
\begin{gathered}
    -\frac{200}{399}q_y^t \Delta q_x^2 + \frac{3}{2660}q_y^t q_x^t\Delta q_x \ge 0 \\
    0 < \Delta q_x \le \frac{9}{4000}q_x
\end{gathered}
\end{equation}

\begin{figure}[H]
    \centering
    \includegraphics[width=\textwidth]{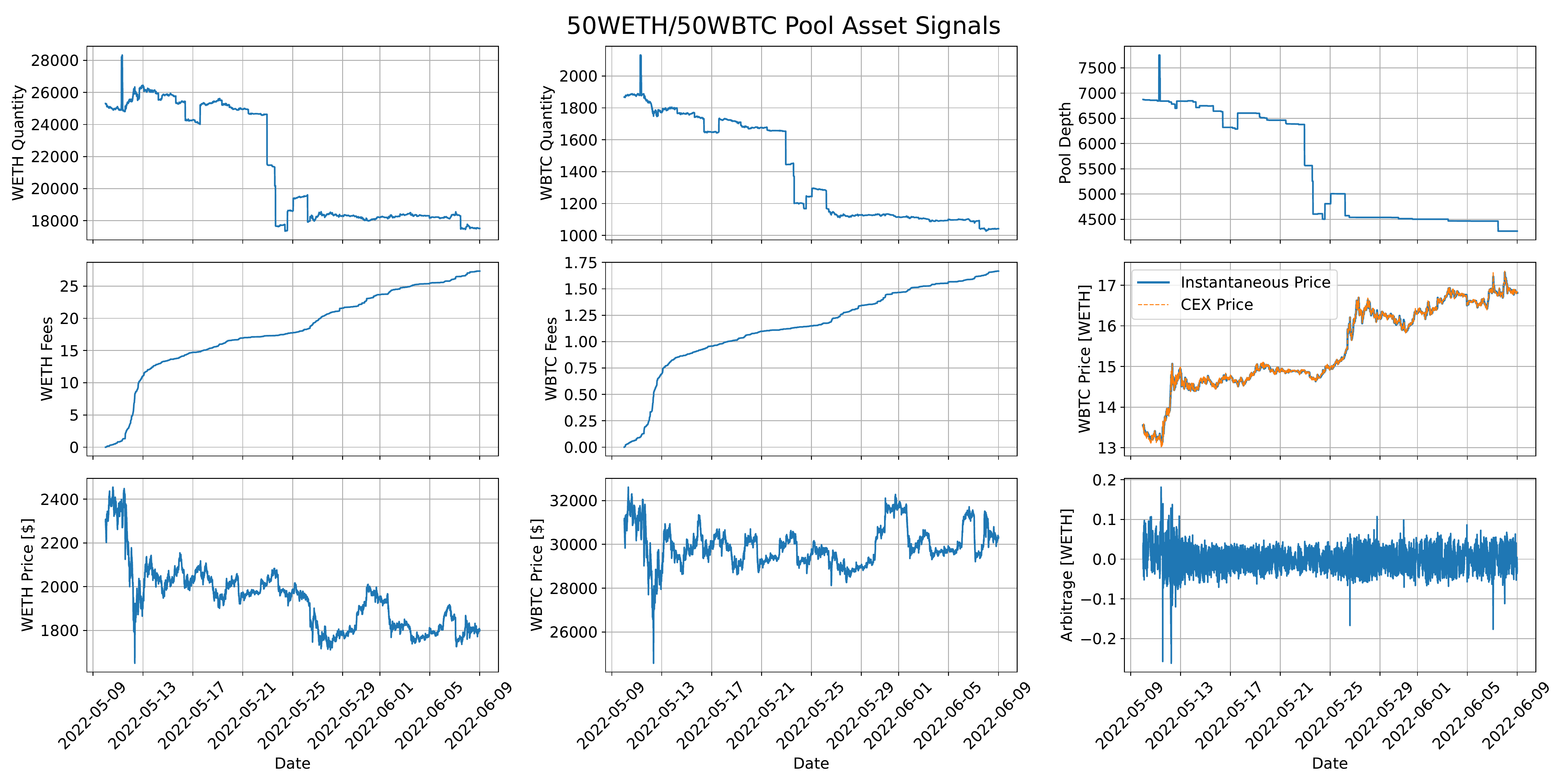}
    \caption{The high correlation between WBTC and WETH quantities ($\rho = 0.98408497$) indicates that the majority of activity in the pool is explained by quotes rather than trades}
    \label{fig:balancer-assets}
\end{figure}

\begin{figure}[H]
    \centering
    \includegraphics[width=\textwidth]{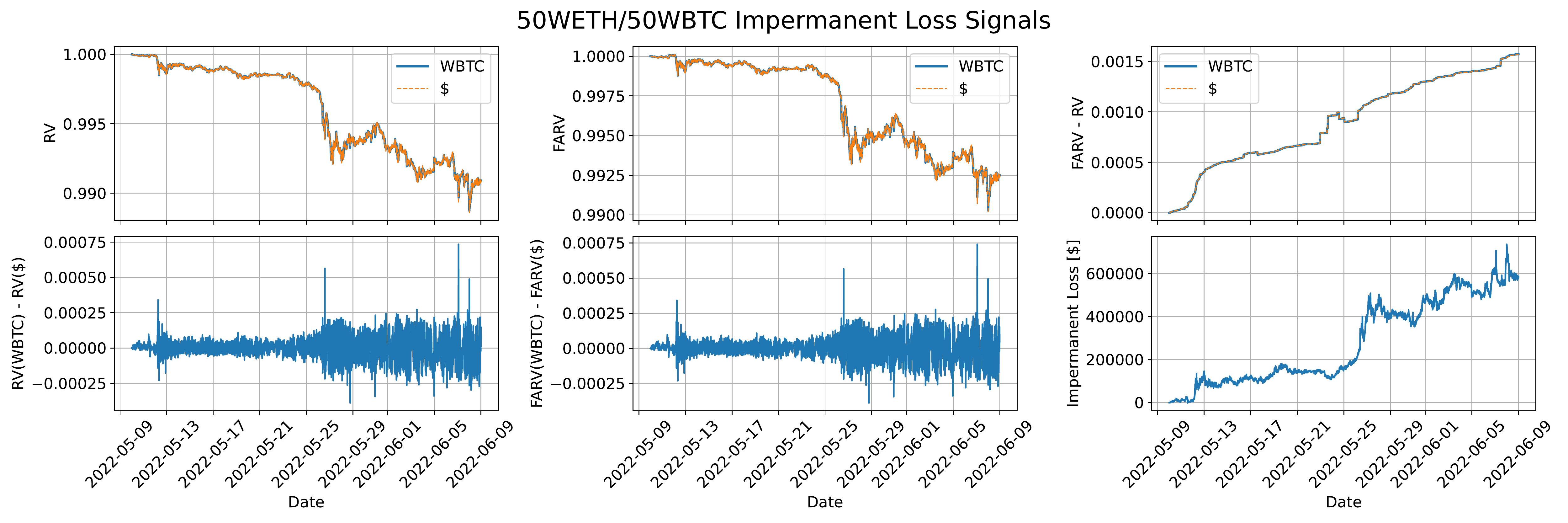}
    \caption{IL is always positive because of the convexity of the AMM function (\cite{permanent-loss}), and FARV shows that a passive LP would have been roughly 1\% wealthier if they had held their assets separate from the pool. FARV and RV currencies do not represent the unit of the signal; rather, they represent the currency with which the signal was generated.}
    \label{fig:balancer-il}
\end{figure}

Over the month period in Figure \ref{fig:balancer-il}, 2472 out of 2622 individual trades would be profitable for an active LP to participate in. That is, ignoring gas fees, an LP participating in single trades would be profitable for roughly 94\% of the trades as defined by the LP profitability condition.

\section{Conclusion}

This paper provides a general framework to understand $N$ asset pool dynamics for any constant product market maker, fee structure and optional concentrated liquidity. We introduce relative value as a more intuitive measure than impermanent loss, and define fee-adjusted relative value as a measure for LP profitability. The framework itself is validated with simulations in UniswapV3 and BalancerV2, where in both cases we observe that the passive pool suffers more from impermanent loss than it gains in fees. However, the LP profitability condition in BalancerV2 indicates that more active LPs have the potential to be profitable over the same period, with 94\% of individual trades fulfilling the LP profitability condition. Unfortunately, the most popular LP strategy is to passively invest, with 30\% of positions on UniswapV3 being held for over a month (\cite{https://doi.org/10.48550/arxiv.2111.09192}). Therefore, Compass Labs recommends LPs to invest more actively, and proposes that DEXs should market make more dynamically too, as opposed to the static pools currently observed.

\bibliographystyle{unsrtnat}
\bibliography{references}  

\begin{thebibliography}{6}
\providecommand{\natexlab}[1]{#1}
\providecommand{\url}[1]{\texttt{#1}}
\expandafter\ifx\csname urlstyle\endcsname\relax
  \providecommand{\doi}[1]{doi: #1}\else
  \providecommand{\doi}{doi: \begingroup \urlstyle{rm}\Url}\fi

\bibitem[Cartea et~al.(2022)Cartea, Drissi, and Monga]{permanent-loss}
Álvaro Cartea, Fayçal Drissi, and Marcello Monga.
\newblock Decentralised finance and automated market making: Predictable loss
  and optimal liquidity provision.
\newblock 2022.
\newblock \doi{http://dx.doi.org/10.2139/ssrn.4273989}.

\bibitem[Loesch et~al.(2021)Loesch, Hindman, Richardson, and
  Welch]{https://doi.org/10.48550/arxiv.2111.09192}
Stefan Loesch, Nate Hindman, Mark~B Richardson, and Nicholas Welch.
\newblock Impermanent loss in uniswap v3, 2021.
\newblock URL \url{https://arxiv.org/abs/2111.09192}.

\bibitem[Bichuch and
  Feinstein(2022)]{https://doi.org/10.48550/arxiv.2210.01227}
Maxim Bichuch and Zachary Feinstein.
\newblock Axioms for automated market makers: A mathematical framework in
  fintech and decentralized finance, 2022.
\newblock URL \url{https://arxiv.org/abs/2210.01227}.

\bibitem[Adams et~al.(2021)Adams, Zinsmeister, Salem, Keefer, and
  Robinson]{uniswapv3}
Hayden Adams, Noah Zinsmeister, Moody Salem, River Keefer, and Dan Robinson.
\newblock Uniswap v3 core.
\newblock 2021.
\newblock URL \url{https://uniswap.org/whitepaper-v3.pdf}.

\bibitem[Angeris et~al.(2019)Angeris, Kao, Chiang, Noyes, and
  Chitra]{https://doi.org/10.48550/arxiv.1911.03380}
Guillermo Angeris, Hsien-Tang Kao, Rei Chiang, Charlie Noyes, and Tarun Chitra.
\newblock An analysis of uniswap markets, 2019.
\newblock URL \url{https://arxiv.org/abs/1911.03380}.

\bibitem[Martinelli and Mushegian(2019)]{balancerv2}
Fernando Martinelli and Nikolai Mushegian.
\newblock A non-custodial portfolio manager, liquidity provider, and price
  sensor.
\newblock 2019.
\newblock URL \url{https://balancer.fi/whitepaper.pdf}.

\end{thebibliography}






\end{document}